\documentclass[10pt,a4paper,twocolumn]{article}
\usepackage{varioref}
\usepackage[dvips]{graphicx}
\usepackage[latin1]{inputenc}

\begin{document}

\vspace{2.5cm}
\begin{center}
{\Large \bf Multistrange Hyperon Production in Pb+Pb collisions at 30, 40, 80 and 158 A$\cdot$GeV}
\end{center} 
\vspace{0.3cm}
{\bf Michael Mitrovski}\hspace{0.1cm}{\bf for the NA49 Collaboration}\footnote [1] {Presented at 25th International School of Nuclear Physics, Erice, Italy} \\ \\ 
Institut f\"ur Kernphysik, August-Euler-Strasse 6, 60486 Frankfurt, Germany\\ \\
{\bf E--mail: Michael.Mitrovski@cern.ch} \\ \\
\noindent
C.~Alt$^{9}$, T.~Anticic$^{21}$, B.~Baatar$^{8}$, D.~Barna$^{4}$,
J.~Bartke$^{6}$,  M.~Behler$^{13}$,
L.~Betev$^{9}$, H.~Bia{\l}\-kowska$^{19}$, A.~Billmeier$^{9}$,
C.~Blume$^{7}$
% ...

\begin{figure}[h!]
\begin{center}
\includegraphics[scale=0.43]{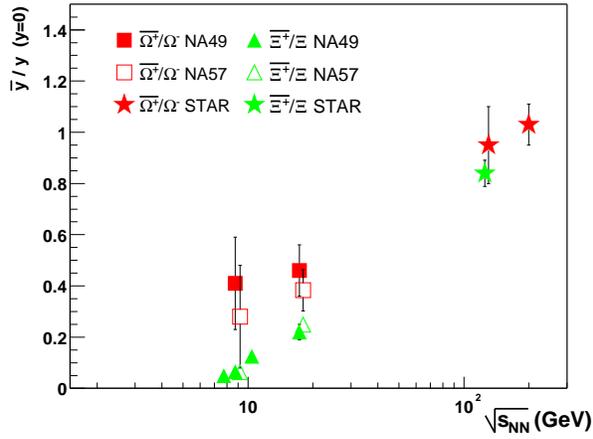}
\end{center}
\caption{The antihyperon/hyperon ($\bar{Y}$$/$$Y$) ratio at midrapidity in the SPS-RHIC energy range.}  \vspace{-0.3cm}
\label{fig:Multistrange}
\end{figure}

\vspace{0.5cm}
\noindent
% ...
$^{4}$KFKI Research Institute for Particle and Nuclear Physics, Budapest, Hungary.\\
% ...
$^{6}$Institute of Nuclear Physics, Cracow, Poland.\\
$^{7}$Gesellschaft f\"{u}r Schwerionenforschung (GSI), Darmstadt, Germany.\\
$^{8}$Joint Institute for Nuclear Research, Dubna, Russia.\\
$^{9}$Fachbereich Physik der Universit\"{a}t, Frankfurt, Germany.\\
% ...
$^{13}$Fachbereich Physik der Universit\"{a}t, Marburg, Germany.\\
% ...
$^{19}$Institute for Nuclear Studies, Warsaw, Poland.\\
% ...
$^{21}$Rudjer Boskovic Institute, Zagreb, Croatia.\\ \\ 
A non-monotonic energy dependence of the $K^{+}$/$\pi^{+}$ ratio with a sharp maximum close to 30 A$\cdot$GeV is observed in central Pb+Pb collisions  \cite{A.2}. Within a statistical model of the early stage \cite{A.3}, this is interpreted as a sign of the phase transition to a QGP, which causes a sharp change in the energy dependence of the strangeness to entropy ratio. This observation naturally motivates us to study the production of multistrange hyperons ($\Xi$, $\Omega$) as a function of the beam energy. \\
Furthermore it was suggested that the kinematic freeze-out of $\Omega$ takes place directly at QGP hadronization. If this is indeed the case, the transverse momentum spectra of the $\Omega$ directly reflect the transverse expansion velocity of a hadronizing QGP \cite{A.5, A.6}. \\   
In this report we show preliminary NA49 results on $\Omega^{-}$ and $\bar{\Omega}^{+}$ production in central Pb+Pb collisions at 40 and 158 A$\cdot$GeV and compare them to measurements of $\Xi^{-}$ and $\bar{\Xi}^{+}$ production in central Pb+Pb collisions at 30, 40, 80 and 158 A$\cdot$GeV. \\ \\
The NA49 detector \cite{A.7} is a large acceptance hadron spectrometer at the CERN SPS, consisting of four TPCs. Two of them, the Vertex TPCs (VTPC), are inside a magnetic field for the determination of particle momenta and charge. The ionisation energy loss (dE/dx) measurements in the two Main TPCs (MTPC), which are outside the magnetic field, are used for mass determination. Central collisions were selected by a trigger using information from a downstream calorimeter (VCAL),  which measures the energy of the projectile spectator nucleons. \\ \\
In Fig.\ref{fig:Multistrange} the NA49 $\bar{\Omega}^{+}$/$\Omega^{-}$ and $\bar{\Xi}^{+}$/$\Xi^{-}$ ratios as a function of the collision energy ($\sqrt{s_{NN}}$) are shown and compared to results of NA57 \cite{A.11,A.12} and STAR \cite{A.13, A.14}. The NA49 and NA57 results measured at the same energies are consistent. The data show a clear increase of the $\bar{\Omega}^{+}$/$\Omega^{-}$ ratio from a value of about 0.4 at SPS energies to about 1 at RHIC energies. The $\bar{\Xi}^{+}$/$\Xi^{-}$ ratio also increases from SPS energies to about 0.8 at RHIC energies.  
\begin{figure}[h!]
\begin{center}
\includegraphics[scale=0.44]{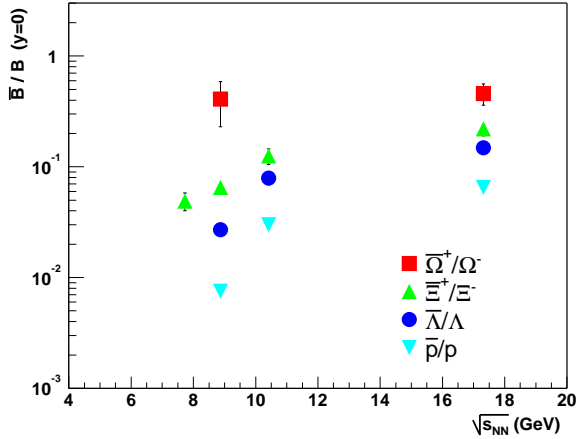}
\end{center}
\caption{The antibaryon/baryon ratio ($\bar{B}$$/$$B$) at midrapidity in the SPS energy range measured by NA49.}  \vspace{-0.3cm}
\label{fig:Baryon}
\end{figure} \\ \\
The midrapidity $\bar{\Omega}^{+}$/$\Omega^{-}$ ratio is estimated to be 0.46 $\pm$ 0.1 and 0.41 $\pm$ 0.18 for central Pb+Pb collisions at 158 and 40 A$\cdot$GeV, respectively. The values for the midrapidity $\bar{\Xi}^{+}$/$\Xi^{-}$ ratio are estimated to be 0.13 $\pm$ 0.02, 0.065 $\pm$ 0.05 and 0.0049 $\pm$ 0.009 for central Pb+Pb collisions at 80, 40 and 30 A$\cdot$GeV, respectively. In Fig.\ref{fig:Baryon} the antibaryon/baryon ratios are shown as a function of the beam energy in the SPS energy domain. In addition to $\bar{\Xi}$/$\Xi$ and $\bar{\Omega}$/$\Omega$ ratios the results on $\bar{\Lambda}$/$\Lambda$ \cite{A.20} and $\bar{p}$/$p$ \cite{A.10} are shown. The energy dependence of $\bar{B}$$/$$B$ ratios gets weaker with increasing strangeness content. 
\begin{figure}[h!]
\begin{center}
\includegraphics[scale=0.43]{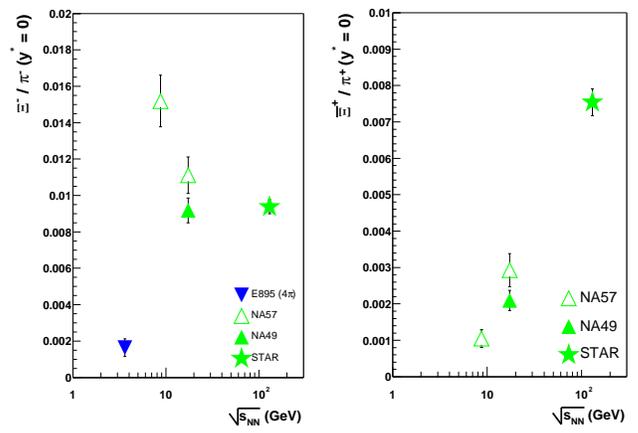}
\end{center}
\caption{Energy dependence of the midrapidity $\Xi^{-}$/$\pi^{-}$ (left) and ${\bar{\Xi}}^{+}$/$\pi^{+}$ (right) ratio in central Pb+Pb and Au+Au collisions.}  \vspace{-0.3cm}
\label{fig:XiPi}
\end{figure} 
\begin{figure}[h!]
\begin{center}
\includegraphics[scale=0.43]{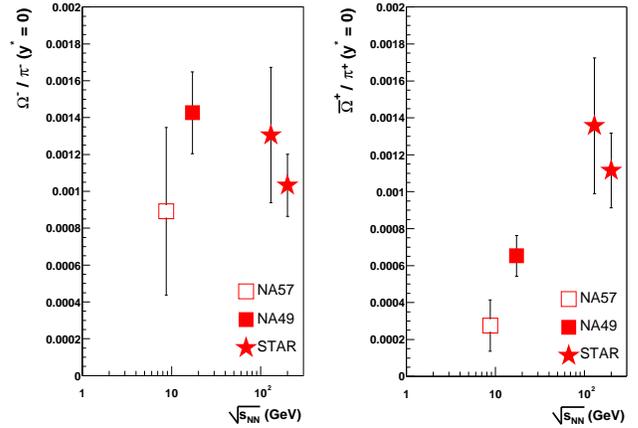}
\end{center}
\caption{Energy dependence of the midrapidity $\Omega^{-}$/$\pi^{-}$ (left) and ${\bar{\Omega}}^{+}$/$\pi^{+}$ (right) ratio in central Pb+Pb and Au+Au collisions.}  \vspace{-0.3cm}
\label{fig:OmegaPi}
\end{figure} \\ \\
Fig.~\ref{fig:XiPi} shows the energy dependence of the midrapidity $\Xi^{-}$/$\pi^{-}$ (left) and ${\bar{\Xi}}^{+}$/$\pi^{+}$ (right) ratio in central Pb+Pb and Au+Au collisions. The $\Xi^{-}$/$\pi^{-}$ ratio suggests that there is a non-monotonic energy dependence at SPS energies. The ${\bar{\Xi}}^{+}$/$\pi^{+}$ ratio increases with energy from SPS to RHIC energies. The $\Omega^{-}$/$\pi^{-}$ ratio shown in Fig.~\ref{fig:OmegaPi} seems to be energy independent, but the ${\bar{\Omega}}^{+}$/$\pi^{+}$ ratio shows again an increase from SPS to RHIC energies. \\ \\
At 158 A$\cdot$GeV, a high statistics data sample of central Pb+Pb collisions is available, which allows us to obtain fully corrected spectra of $\Omega^{-}$ and $\bar{\Omega^{+}}$. The transverse mass spectra are fitted by an exponential function :
\begin{equation}
\frac{1}{m_{t}} \frac{d^{2}N}{dm_{t}dy} = C \cdot e^{- m_{t}/T}, 
\end{equation}
where the fit parameters are a normalization factor \textit{C} and the inverse slope parameter \textit{T}. The slope parameter is similar for $\Omega^{-}$ and $\bar{\Omega}^{+}$ : $T$($\Omega^{-}$) = 276 $\pm$ 23 MeV and $T$($\bar{\Omega}^{+}$) = 285 $\pm$ 39 MeV \cite{A.10}. Our values agree with those measured by the NA57 collaboration ($T$($\Omega^{-}$) = 280 $\pm$ 16 MeV and $T$($\bar{\Omega}^{+}$) = 324 $\pm$ 29 MeV) \cite{A.15}. The large acceptance of the NA49 experiment allows us to measure the $\Omega^{-}$ ($\bar{\Omega}^{+}$) spectra in a large rapidity interval. The rapidity distributions for $\Omega^{-}$ and  $\bar{\Omega}^{+}$ are obtained by extrapolating $p_{t}$ spectra using the exponential. Both y-spectra were fitted by a Gaussian. The width of the $\Omega^{-}$ distribution ($\sigma$($\Omega^{-}$) = 1.0 $\pm$ 0.2) seems to be larger than the one of the $\bar{\Omega}^{+}$ ($\sigma$($\Omega^{-}$)  = 0.7 $\pm$ 0.1). Mean multiplicities in full phase-space were estimated as integrals over measured points corrected for the missing rapidity coverage using the Gaussian parametrisations. The resulting yields are $<$$\Omega^{-}$$>$ = 0.47 $\pm$ 0.07 and $<$$\bar{\Omega}^{+}$$>$ = 0.15 $\pm$ 0.02, where the errors are statistical only. 
\section*{Acknowledgments} \vspace{-0.3cm}
This work was supported by the US Department of Energy
Grant DE-FG03-97ER41020/A000,
the Bundesministerium fur Bildung und Forschung, Germany, 
the Polish State Committee for Scientific Research (2 P03B 130 23, SPB/CERN/P-03/Dz 446/2002-2004, 2 P03B 02418, 2 P03B 04123), 
the Hungarian Scientific Research Foundation (T032648, T032293, T043514),
the Hungarian National Science Foundation, OTKA, (F034707),
and the Polish-German Foundation.

\end{document}